\documentclass[aps,preprint]{revtex4}%
\usepackage{amsfonts}
\usepackage{amsmath}
\usepackage{amssymb}
\usepackage{graphics}
\usepackage{graphicx}%
\providecommand{\U}[1]{\protect\rule{.1in}{.1in}}

\begin{document}
\title{Exploring Weak Turbulence of Phonon and Magnon Beams in Magneto-Acoustic Ultrathin Films }
\author{Vladimir L. Safonov, Derek A. Bas,  Andrew Franson, Piyush J. Shah, Michael E. McConney, Michael Newburger,
Michael R. Page}
\affiliation{Air Force Research Laboratory, Materials and Manufacturing Directorate,
Wright-Patterson Air Force Base, OH 45433.}

\begin{abstract}
This study presents a simple theoretical model describing narrow envelope surface acoustic waves (phonons) and spin waves (magnons) in an ultrathin ferromagnetic film. Based on the general principles of weak wave turbulence, the model considers interactions between beams of an ideal phonon gas and a weakly non-ideal magnon gas, which represent magnetoacoustic oscillations in the system. Equations for the wave envelopes of phonons and magnons, along with their harmonics, are derived, incorporating nonlinear effects from three- and four-particle interactions. In the general non-resonant case, linear stationary envelope simulations are sufficient. These clarify the experimentally observed angular dependence of the transmitted acoustic signal with respect to the orientation of the magnetic field. The study highlights increased energy losses associated with enhanced magnetoacoustic coupling. Given the broad interdisciplinary interest in weak turbulence phenomena within condensed matter physics and nonlinear wave dynamics, our model offers significant predictive capabilities and greatly simplifies calculations of quasiparticle beam interactions.

\end{abstract}
\maketitle

\section{Introduction}

The study of surface acoustic waves in ultrathin magnetic films is of interest for fundamental and applied science [1]-[6]. On the one hand, we have a linear wave system of Rayleigh surface acoustic waves (phonons). On the other hand, this system consists of spin waves (magnons), in which small deviations of magnetization lead to nonlinear effects. It is obvious that the magnetoelastic interaction, which connects magnetic deviations with elastic deformations, is most pronounced at and near the intersection of the phonon and magnon spectra. Linear excitation of spin waves by surface sound near the intersection of their spectra is usually called acoustically-driven
ferromagnetic resonance. It should be noted, however, that magnetoelastic effects are experimentally observed for surface acoustic waves when the frequencies of phonons and magnons at a given wave vector differ significantly. We clarify this effect in this paper. Thus, we have a rich nonlinear wave system that can be excited by both acoustic and radio frequency fields and controlled by an external magnetic field.

The most common method for describing interacting elastic and spin waves is direct computer modeling. The sample is divided into many cells. Each cell utilizes the Landau-Lifshitz-Gilbert equation, which determines the local oscillations of magnetization in response to effective fields from both the magnetic and elastic systems. Local elastic oscillations of the film are described by oscillations of the strain tensor, considering the stresses from the magnetic system. This approach may seem quite comprehensive, but this is a misconception, since the problem necessarily considers phenomenological attenuation, in which the processes of energy redistribution in the thermostat are already averaged, and thus, the microscopic dynamics become truncated. With this approach, by fitting several phenomenological parameters to the problem, it is potentially possible to align the simulation results with the experimental data. However, the predictive part of such a theory is minimal. The disadvantages of such modeling include high requirements for computing power and computing time. The whole point is that the dynamics of all degrees of freedom of the system are calculated, most of which do not play a noticeable role in the process of interaction of phonons and magnons in a narrow region of the spectrum.

Since the most interesting physical processes occur in the vicinity of the intersection point of the elastic and magnetic oscillation spectra, the role of a vast number of degrees of freedom from other spectral regions in the problem can be neglected (their role is taken into account in the relaxation parameters). Here we develop the theory of narrow wave packets (envelopes) describing beams of quasiparticles (phonons and magnons). This approach significantly simplifies theoretical calculations and has a high predictive potential. The concept of envelopes was developed in the theory of weak wave turbulence \cite{ZMR},\cite{ZLF}. Our contribution to the development of this concept is that we study the weak turbulence of two interacting flows of waves of different types with different spectra. In this case, in the general case, the flow of phonons linearly non-resonantly excites the flow of magnons and also non-resonantly receives feedback.

In this paper, we derive a system of fundamental equations for describing wave packets of phonons and magnons moving in the same direction, taking into account three- and four-particle interactions, as well as for wave packets of the second and third elastic harmonics. This results in a rich nonlinear system with various solutions depending on many external and internal parameters. These equations can be used to analyze stationary flows (beams) of phonons and magnons in an ultrathin ferromagnetic film. In the nonresonant case, when the amplitudes of the excited magnons are noticeably weakened, nonlinear interactions can be neglected at first. We perform simulations of linear stationary flows of phonons and magnons depending on the magnitude of the magnetoelastic interaction, the magnetic field, the film size, and the frequency of the excited phonons. The paper concludes with a discussion. All the numerous but necessary for the calculations details are included in the supplemental material.

\section{Magnons and Phonons}

We will describe the motions of the magnetization vector $\boldsymbol{M}$
using $\boldsymbol{k}$-space and classical Hamiltonian formalism (all details
are given in the supplement) using complex variables $c_{\boldsymbol{k}}%
^{\ast}$ and $c_{\boldsymbol{k}}$, which are analogs of the Bose creation and
annihilation operators of magnons in quantum mechanics \cite{ZLF}-\cite{VS}.
The Hamiltonian of magnons can be written as
\begin{equation}
\mathcal{H}_{m}/\hbar=\sum\limits_{\boldsymbol{k}}\omega_{m\boldsymbol{k}%
}c_{\boldsymbol{k}}^{\ast}c_{\boldsymbol{k}}+\frac{1}{2}\sum
\limits_{\boldsymbol{k}_{1},\boldsymbol{k}_{2},\boldsymbol{k}_{3}%
,\boldsymbol{k}_{4}}\Phi_{4m}(\boldsymbol{k}_{1},\boldsymbol{k}_{2}%
;\boldsymbol{k}_{3},\boldsymbol{k}_{4})c_{\boldsymbol{k}_{1}}^{\ast
}c_{\boldsymbol{k}_{2}}^{\ast}c_{\boldsymbol{k}_{3}}c_{\boldsymbol{k}_{4}%
}\Delta(\boldsymbol{k}_{1}+\boldsymbol{k}_{2}-\boldsymbol{k}_{3}%
-\boldsymbol{k}_{4}), \label{universal H}%
\end{equation}
where the first term describes an ideal gas of magnons with the spectrum
$\omega_{m\boldsymbol{k}}$ and the second term describes the nonlinear
scattering with the amplitude $\Phi_{4m}(\boldsymbol{k}_{1},\boldsymbol{k}%
_{2};\boldsymbol{k}_{3},\boldsymbol{k}_{4})$, $\Delta(\boldsymbol{k})$ is the
Kronecker delta function.\ The Plank's constant $\hbar$ is used to match with
quantum mechanics and as a dimensional parameter. The Hamiltonian
(\ref{universal H}) models the weak turbulence of magnons in a magnetic film.

The dynamic equation for (\ref{universal H}) can be written in the following
form:
\begin{equation}
i\left(  \frac{\partial}{\partial t}+\eta_{m\boldsymbol{k}}\right)
c_{\boldsymbol{k}}-\omega_{m\boldsymbol{k}}c_{\boldsymbol{k}}=\sum
\limits_{\boldsymbol{k}_{2},\boldsymbol{k}_{3},\boldsymbol{k}_{4}}\Phi
_{4m}(\boldsymbol{k},\boldsymbol{k}_{2};\boldsymbol{k}_{3},\boldsymbol{k}%
_{4})c_{\boldsymbol{k}_{2}}^{\ast}c_{\boldsymbol{k}_{3}}c_{\boldsymbol{k}_{4}%
}\Delta(\boldsymbol{k}+\boldsymbol{k}_{2}-\boldsymbol{k}_{3}-\boldsymbol{k}%
_{4}), \label{Main Ham eq}%
\end{equation}
where the energy loss term with the phenomenological relaxation $\eta
_{m\boldsymbol{k}}$ is included.

\subsection{Magnon Wave Envelop}

Consider a narrow wave packet $\boldsymbol{k}=\boldsymbol{k}_{e}%
+\boldsymbol{q}$ with the central beam wave vector $\boldsymbol{k}_{e}%
=\{k_{e},0\}$ and $\left\vert \boldsymbol{q}\right\vert \ll\left\vert
\boldsymbol{k}_{e}\right\vert $. Then the frequency can be represented as an
expansion
\begin{equation}
\omega_{mk}\simeq\omega_{mk_{e}}+v_{g}q_{z}\boldsymbol{+}\frac{v_{g}}{2k_{e}%
}q_{x}^{2}+\frac{\omega^{\prime\prime}}{2}q_{z}^{2}. \label{freq exp final}%
\end{equation}
Here $\boldsymbol{v}_{g}=\left(  \partial\omega_{mk}/\partial k\right)
_{\boldsymbol{k}=\boldsymbol{k}_{e}}$ is the magnon group velocity.and
$\omega^{\prime\prime}\equiv\left(  \partial^{2}\omega_{mk}/\partial
k^{2}\right)  _{\boldsymbol{k}=\boldsymbol{k}_{e}}$.

Let us assume that the magnon envelope is linearly excited by an external wave
with wave vector $\boldsymbol{k}_{e}$ and frequency $\Omega_{k_{e}}$. Then the
complex magnon amplitude will oscillate with frequency $\Omega_{k_{e}}$:
$c_{k_{e}}(t)\propto\exp\left(  -i\Omega_{k_{e}}t\right)  $. We can introduce
a slow variable $\psi_{m}(\boldsymbol{r,t})$ which describe the local dynamics
of the envelop:%

\begin{equation}
c_{\boldsymbol{k}}=\frac{1}{L_{x}L_{z}}\int_{0}^{L_{x}}dx\int_{0}^{Lz}%
dz\ \psi_{m}(\boldsymbol{r},t)\exp(-i\Omega_{k_{e}}t-i\boldsymbol{qr}).
\label{intro psi}%
\end{equation}
Having carried out a series of simple transformations, we obtain the magnon
envelop equation
\begin{align}
&  i\frac{\partial\psi_{m}(\boldsymbol{r},t)}{\partial t}+\left(
i\eta_{m\boldsymbol{k}_{e}}+\Delta\omega_{k_{e}}\right)  \psi_{m}%
(\boldsymbol{r},t)+iv_{g}\frac{\partial}{\partial z}\psi_{m}(\boldsymbol{r}%
,t)\label{psi env}\\
&  +\frac{\omega^{\prime\prime}}{2}\frac{\partial^{2}}{\partial z^{2}}\psi
_{m}(\boldsymbol{r},t)+\frac{v_{g}}{2k_{e}}\frac{\partial^{2}}{\partial x^{2}%
}\psi_{m}(\boldsymbol{r},t)-\Phi_{4m,e}\left\vert \psi_{m}(\boldsymbol{r,t}%
)\right\vert ^{2}\psi_{m}(\boldsymbol{r},t)=0,\nonumber
\end{align}
where $\Delta\omega_{k_{e}}=\Omega_{k_{e}}-\omega_{mk_{e}}$and $\Phi
_{4m,e}\equiv\Phi_{4m}(\boldsymbol{k}_{e},\boldsymbol{k}_{e};\boldsymbol{k}%
_{e},\boldsymbol{k}_{e})$. This equation with $\Delta\omega_{k_{e}}=0$,
without taking into account the dissipative term, is encountered in many areas
of nonlinear physics \cite{NLSEbook},\cite{NGO},\cite{Falk}. It is called the
nonlinear Schr\"{o}dinger equation or the Gross-Pitaevsky equation. The
amplitude $\Phi_{4m,b}$ can be both positive (repulsive interaction) and
negative (attraction). In the second case, the intense flow of waves narrows,
and nonlinear self-focusing occurs.

In the case when $\Delta\omega_{k_{e}}\neq0$ magnon envelope is excited
non-resonantly, on the wing of the resonance lineshape. In the general case,
we have this option when a surface acoustic wave with wave vector
$\boldsymbol{k}_{e}$ passes through an ultra-thin magnetic film and
non-resonantly excites magnetic oscillations with the same wave vector but
different frequencies.

\subsection{Phonon and Magnon Wave Envelops}

Let us now consider a phonon envelope. For simplicity, we will consider only
longitudinal phonons with their polarization along the direction of motion.
These waves model Rayleigh waves in the plane \cite{Dreher}. The phonon
Hamiltonian contains the quadratic form only
\begin{equation}
\mathcal{H}_{ph}/\hbar=\sum_{\boldsymbol{k}}v_{s}kb_{\boldsymbol{k}}^{\ast
}b_{\boldsymbol{k}}, \label{ph-Ham}%
\end{equation}
where $v_{s}$ is the sound velocity and $b_{\boldsymbol{k}}^{\ast},$
$b_{\boldsymbol{k}}$ are the complex phonon amplitudes. The dynamic equation
for $b_{\boldsymbol{k}}$ is similar to Eq.(\ref{Main Ham eq}) without nonlinearity.

Magnetoelastic interaction between magnetization and elastic deformations at
the point of the intersection of their spectra leads to processes of
interaction of magnons and phonons.

Let us derive equations for envelopes of phonons and magnons resonantly
interacting. Only the envelopes moving along the $z$ axis will be considered.
In this case, the following terms must be added to the model magnon
Hamiltonian (\ref{universal H}) and phonon Hamiltonian (\ref{ph-Ham})%
\begin{align}
\mathcal{H}_{mph}/\hbar &  =\left(  \digamma_{\boldsymbol{k}_{e}%
}c_{\boldsymbol{k}_{e}}^{\ast}b_{\boldsymbol{k}_{e}}+\text{c.c.}\right)
+\Phi_{2m2ph}(\boldsymbol{k}_{e},\boldsymbol{k}_{e};\boldsymbol{k}%
_{e},\boldsymbol{k}_{e})c_{\boldsymbol{k}_{e}}^{\ast}c_{\boldsymbol{k}_{e}%
}b_{\boldsymbol{k}_{e}}^{\ast}b_{\boldsymbol{k}_{e}}\label{m-e interaction}\\
&  +\left[  \Phi_{pump}(\boldsymbol{k}_{e},\boldsymbol{k}_{e};\boldsymbol{k}%
_{e},\boldsymbol{k}_{e})c_{\boldsymbol{k}_{e}}c_{\boldsymbol{k}_{e}%
}b_{\boldsymbol{k}_{e}}^{\ast}b_{\boldsymbol{k}_{e}}^{\ast}+\text{c.c.}\right]
\nonumber\\
&  +\left[  \Phi_{3mph}(\boldsymbol{k}_{e},\boldsymbol{k}_{e},\boldsymbol{k}%
_{e};\boldsymbol{k}_{e})c_{\boldsymbol{k}_{e}}^{\ast}c_{\boldsymbol{k}_{e}%
}c_{\boldsymbol{k}_{e}}b_{\boldsymbol{k}_{e}}^{\ast}+\text{c.c.}\right]
\nonumber\\
&  +\left[  \Psi_{2mph2}(\boldsymbol{k}_{e},\boldsymbol{k}_{e};2\boldsymbol{k}%
_{e})c_{\boldsymbol{k}_{e}}c_{\boldsymbol{k}_{e}}b_{2\boldsymbol{k}_{e}}%
^{\ast}+\text{c.c.}\right] \nonumber\\
&  +\left[  \Phi_{3mph3}(\boldsymbol{k}_{e},\boldsymbol{k}_{e},\boldsymbol{k}%
_{e};3\boldsymbol{k}_{e})c_{\boldsymbol{k}_{e}}c_{\boldsymbol{k}_{e}%
}c_{\boldsymbol{k}_{e}}b_{3\boldsymbol{k}_{e}}^{\ast}+\text{c.c.}\right]
.\nonumber
\end{align}
The first term with amplitude $U_{\boldsymbol{k}_{e}}$ describes the linear
excitation of the magnon by a resonant phonon and the inverse process. The
nonlinear addition to this process is written in the form of a second, third,
and fourth terms with amplitude $\Phi_{2m2ph}$, $\Phi_{pump}$ and $\Phi
_{3mph}$, correspondingly. The fifth term with amplitude $\Psi_{2mph2}$
describes the transformation of two magnons into a phonon with doubled
frequency and doubled wave vector (and the reverse process). The last term
with $\Phi_{3mph3}$\ is the transformation of three magnons into a phonon with
tripled frequency and tripled wave vector (and the reverse process).

The resulting picture can be described by four envelopes as follows. If we
initially excite an envelop $\psi_{ph}$\ of phonons along $z$ axis with a
frequency $v_{s}k_{e}$, then due to linear interaction, this phonon flow
non-resonantly excites envelop $\psi_{m}$ of magnons on the wing of their
resonance lineshape in the same direction with frequency $v_{s}k_{e}$. Then
magnon envelop excites second $\psi_{ph2}$ and third $\psi_{ph3}$\ sound
harmonics.%
\begin{align}
&  i\frac{\partial\psi_{ph}}{\partial t}+i\eta_{ph}\psi_{ph}+iv_{s}%
\frac{\partial}{\partial z}\psi_{ph}+\frac{v_{s}}{2k_{e}}\frac{\partial^{2}%
}{\partial x^{2}}\psi_{ph}-\digamma_{\boldsymbol{k}_{e}}^{\ast}\psi
_{m}\label{ph envelop}\\
&  -\Phi_{2m2ph}\left\vert \psi_{m}\right\vert ^{2}\psi_{ph}-2\Phi_{pump}%
\psi_{m}^{2}\psi_{ph}^{\ast}-\Phi_{3mph}\left\vert \psi_{m}\right\vert
^{2}\psi_{m}=0.\nonumber
\end{align}%
\begin{align}
&  i\frac{\partial\psi_{m}}{\partial t}+\left(  i\eta_{m}+v_{s}k_{e}%
-\omega_{mk_{e}}\right)  \psi_{m}+iv_{g}\frac{\partial}{\partial z}\psi
_{m}+\frac{\omega^{\prime\prime}}{2}\frac{\partial^{2}}{\partial z^{2}}%
\psi_{m}+\frac{v_{g}}{2k_{e}}\frac{\partial^{2}}{\partial x^{2}}\psi_{m}%
-\Phi_{4m}\left\vert \psi_{m}\right\vert ^{2}\psi_{m}\label{magnon envelop}\\
&  -\digamma_{\boldsymbol{k}_{e}}\psi_{ph}-\Phi_{2m2ph}\left\vert \psi
_{ph}\right\vert ^{2}\psi_{m}-\Phi_{pump}\psi_{ph}^{2}\psi_{m}^{\ast}%
-\Phi_{3mph}\psi_{m}^{2}\psi_{ph}^{\ast}\nonumber\\
&  -2\Phi_{3mph}^{\ast}\left\vert \psi_{m}\right\vert ^{2}\psi_{ph}%
-2\Psi_{2mph2}^{\ast}\psi_{m}^{\ast}\psi_{ph2}-3\Phi_{3mph3}^{\ast}(\psi
_{m}^{\ast})^{2}\psi_{ph3}=0.\nonumber
\end{align}%
\begin{equation}
i\frac{\partial\psi_{ph2}}{\partial t}+i\eta_{ph2}\psi_{ph2}+iv_{s}%
\frac{\partial}{\partial z}\psi_{ph2}+\frac{v_{s}}{4k_{e}}\frac{\partial^{2}%
}{\partial x^{2}}\psi_{ph2}-\Psi_{2mph2}\psi_{m}^{2}=0. \label{2ph envelop}%
\end{equation}%
\begin{equation}
i\frac{\partial\psi_{ph3}}{\partial t}+i\eta_{ph3}\psi_{ph3}+iv_{s}%
\frac{\partial}{\partial z}\psi_{ph3}+\frac{v_{s}}{6k_{e}}\frac{\partial^{2}%
}{\partial x^{2}}\psi_{ph3}-\Phi_{3mph3}\psi_{m}^{3}=0. \label{3ph envelop}%
\end{equation}
Note that these equations have a general form, and the coefficients of these
equations depend on the specific model of magnetic, elastic and magnetoelastic
properties of the sample.

\section{Model of Ultrathin Ferromagnetic Film}

Consider an ultra-thin ferromagnetic film with a thickness of $\tau$ and
dimensions of $0\leq z\leq L_{z}$, $-L_{x}/2\leq x\leq L_{x}/2$. The
equilibrium is supposed to be a uniformly magnetized state in which the
magnetization $\boldsymbol{M}(z,x)$ is oriented in the $(z,x)$ plane along an
equilibrium axis $z_{0}$ which is deviated from the $z$ axis by the angle
$\phi_{0}$. The magnetic energy of the film contains exchange energy,
uniaxial anisotropy, Zeeman energy, and demagnetization energy. The changes in
magnetization $\boldsymbol{M}(z,x)$ and other physical parameters across the
film thickness are neglected. We provide all the details of the description of
the oscillations of the magnetic and elastic subsystems in the supplementary
text to this article.

Let us consider the specifics of the magnon spectrum, which can be represented
as:
\begin{align}
\omega_{m\boldsymbol{k}}  &  =\gamma\left[  \alpha_{E}k^{2}+H\cos(\phi
_{H}-\phi_{0})+H_{K}\cos2\phi_{0}+4\pi M_{s}G(k\tau)\right]  ^{1/2}%
\label{sw spectrum}\\
&  \times\left[  \alpha_{E}k^{2}+H\cos(\phi_{H}-\phi_{0})+H_{K}\cos^{2}%
\phi_{0}+4\pi M_{s}[1-G(k\tau)]\left(  \frac{k_{x}\cos\phi_{0}-k_{z}\sin
\phi_{0}}{k}\right)  ^{2}\right]  ^{1/2}.\nonumber
\end{align}
Here $\alpha_{E}$ is the exchange constant, $H$ is the external magnetic field
rotated by an angle $\phi_{H}$ from the $z$-axis, $H_{K}$ is the uniaxial
anisotropy field along $z$-axis, $G(\xi)=[1-\exp(-\xi)]/\xi$. We see that the
spectrum of magnons (\ref{sw spectrum}) has non-trivial dependences on the
angles $\phi_{H}$, $\phi_{0}$ and on the wave vector $\boldsymbol{k}%
=(k_{z},k_{x})$. Below, we will consider the beam of phonons and magnons in the
direction of the $z$-axis. The anisotropy field for simplicity will be
neglected ($H_{K}=0$ Oe, $\phi_{0}=\phi_{H}$).

Figures 1a-d show the magnon and phonon spectra at $\phi_{H}=0$ . To calculate
these spectra, the following Ni film parameters were used in
Eq.(\ref{sw spectrum}): $\boldsymbol{k}=(k,0)$, $\tau=70$ nm, $v_{s}%
=3.5\times10^{5}$ cm/s, $\alpha_{E}=8\times10^{-7}$ erg/cm, $M_{s}=484$ Gs,
$\gamma=3.05\times10^{6}$ Hz/Oe. We see that the magnon spectrum has a
negative group velocity at small $k$ (see Fig. 1c), and then it becomes zero at
the point indicated in Fig. 1ao. At larger $k$ we have $v_{g}>0$. Figure 1e
shows 3D phonon and magnon spectra and their intersection. Figure 1f
demonstrate the surface of the phonon and magnon spectra intersection.
%
\begin{figure}[ptbh]
\includegraphics[width=11cm]{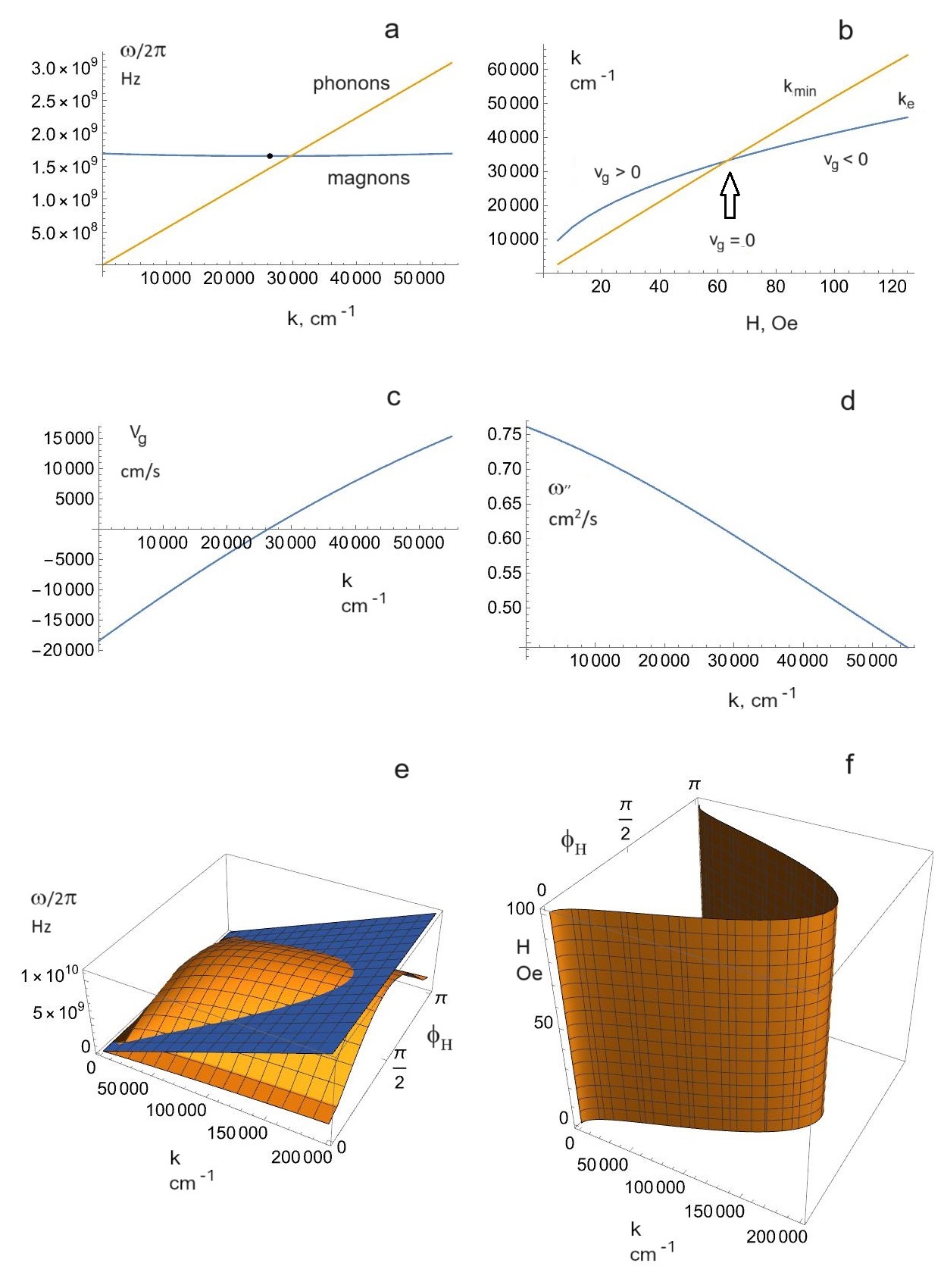}\caption{ Characteristics of
phonon and magnon spectra. a) Spectra of phonons $v_{s}k$ and magnons
$\omega_{m\boldsymbol{k}}$ at $\phi_{H}=0$, $H=50$ Oe. The black dot indicates
the minimum frequency of magnons at $k_{min}$ when their group velocity
$v_{g}=0$. b) Magnetic field dependence of $k_{min}$ when $v_{g}=0$ and of
$k_{e}$, wave vector magnitude of spectra intersection. c) Wave-vector
dependence of the magnon group velocity. d) Wave-vector dependence of the
second derivatives on $k$ for the magnon spectrum. e) 3D representation of
magnon (yellow surface) and phonon (blue surface) spectra showing the spectral
intersection curve. f) The surface describes the intersection points of the
spectra of phonons and magnons,s depending on the magnitude and angle of
rotation of the magnetic field and the wave vector. }%
\end{figure}

From Fig. 1e, f, it is evident that the intersection point of the phonon and
magnon spectra, at which the magnetoelastic interaction is maximal, has a
strong angular dependence on the magnetic field direction. For this reason, a
beam of phonons excited at one frequency non-resonantly excites a beam of
magnons moving in the same direction.

\section{Linear Beams}

Here, we consider the most common case in surface acoustic wave experiments, when the frequency of excited phonons $v_s k_e$
 differs significantly from the frequency $\omega_{mk_e}$ of magnons with the same wave vector $\boldsymbol{k}_e$. In this case, magnons are excited non-resonantly on their lineshape wing, and magnon amplitudes are weakened by the parameter $\eta_m/|\omega_{mk_e}-v_s k_e|\ll 1$. 
Then, nonlinear interactions will be noticeably weakened and can be neglected for simplicity. Thus, we will limit ourselves to the linear interaction of envelopes without nonlinear excitation of the envelopes of the second and third harmonics. We will consider only stationary flows of quasiparticles, which we will call beams. The linear beams of the system (\ref{ph envelop}) and
(\ref{magnon envelop}) are reduced to the following equations, written for
convenience in dimensionless coordinates $X=k_{e}x,$ and $Z=k_{e}z$:
\begin{equation}
i\frac{\eta_{ph}}{v_{s}k_{e}}\psi_{ph}+i\frac{\partial}{\partial Z}\psi
_{ph}+\frac{1}{2}\frac{\partial^{2}}{\partial X^{2}}\psi_{ph}-\frac
{\digamma_{\boldsymbol{k}_{e}}^{\ast}}{v_{s}k_{e}}\psi_{m}=0, \label{Lin mag}%
\end{equation}%
\begin{equation}
\left(  i\frac{\eta_{m}}{v_{s}k_{e}}+1-\frac{\omega_{mk_{e}}}{v_{s}k_{e}%
}\right)  \psi_{m}+i\frac{v_{g}}{v_{s}}\frac{\partial}{\partial Z}\psi
_{m}+\frac{\omega^{\prime\prime}k_{e}}{2v_{s}}\frac{\partial^{2}}{\partial
Z^{2}}\psi_{m}+\frac{1}{2}\frac{v_{g}}{v_{s}}\frac{\partial^{2}}{\partial
X^{2}}\psi_{m}-\frac{\digamma_{\boldsymbol{k}_{e}}}{v_{s}k_{e}}\psi_{ph}=0.
\label{Lin ph}%
\end{equation}

We have calculated the phonon (\ref{Lin mag}) and magnon (\ref{Lin ph}) beams
in a reasonably wide parameter range using Wolfram Mathematica and the finite
element method. Without significant loss of generality, we present the
Ni ultrathin magnetic film calculation results. In addition, we consider
1) a linear emitter of acoustic waves with a vertical size of $10$ $\mu$m
installed on the left in the center at the edge of the film and 2) an
acoustic receiver of the same size installed on the right edge. The
amplitude of the excited bim phonons was assumed to be $\psi_{ph}=1$. The
transmitted signal was calculated using the formula

\begin{equation}
S_{\text{trans}}(\phi_{H})=20\log_{10}\int_{-X_{R}/2}^{X_{R}/2}\left\vert
\psi_{ph}\right\vert dX.\label{trans}%
\end{equation}
The problem was subject to boundary conditions for the magnetic system:
$\psi_{m}=0$ at the left, lower, and upper edges of the magnetic film. We
assumed that $\eta_{ph}/v_{s}k_{e}=\eta_{m}/v_{s}k_{e}=10^{-4}$. The
quantities $\omega_{mk_{e}}$, $v_{g}$ and $\omega^{\prime\prime}$ were
calculated from the spectrum (\ref{sw spectrum}). The amplitude of the linear
magnetoelastic interaction (its derivation is given in the supplement) is: %
\begin{equation}
\digamma_{k}=-\frac{iB_{1}}{2}\sqrt{\frac{\gamma k}{\rho M_{s}v_{s}}}\left(
U_{\boldsymbol{k}}+V_{\boldsymbol{k}}\right)  \sin2\phi_{0}%
,\label{lin amplitude}%
\end{equation}
where $B_{1}$is the magnetoelastic constant ($\approx10^{8}$ erg/cm$^{3}$ for
Ni \ \cite{NiB1}), $\rho$ is the density ($8.9$ g/cm$^{3}$ for Ni),
coefficients $U_{\boldsymbol{k}}$ and $V_{\boldsymbol{k}}$ are given in the Supplement.

Let us first consider how the picture of the beams of phonons and magnons
depends on the value of the magnetoelastic constant $B_{1}$. Figure 2 shows
the beams of phonons and magnons and the angular dependence of the
transmission signal in polar coordinates when the angle $\phi_{H}$ changes
from zero to $\pi/2$.
%
\begin{figure}[ptbh]
\includegraphics[width=12cm]{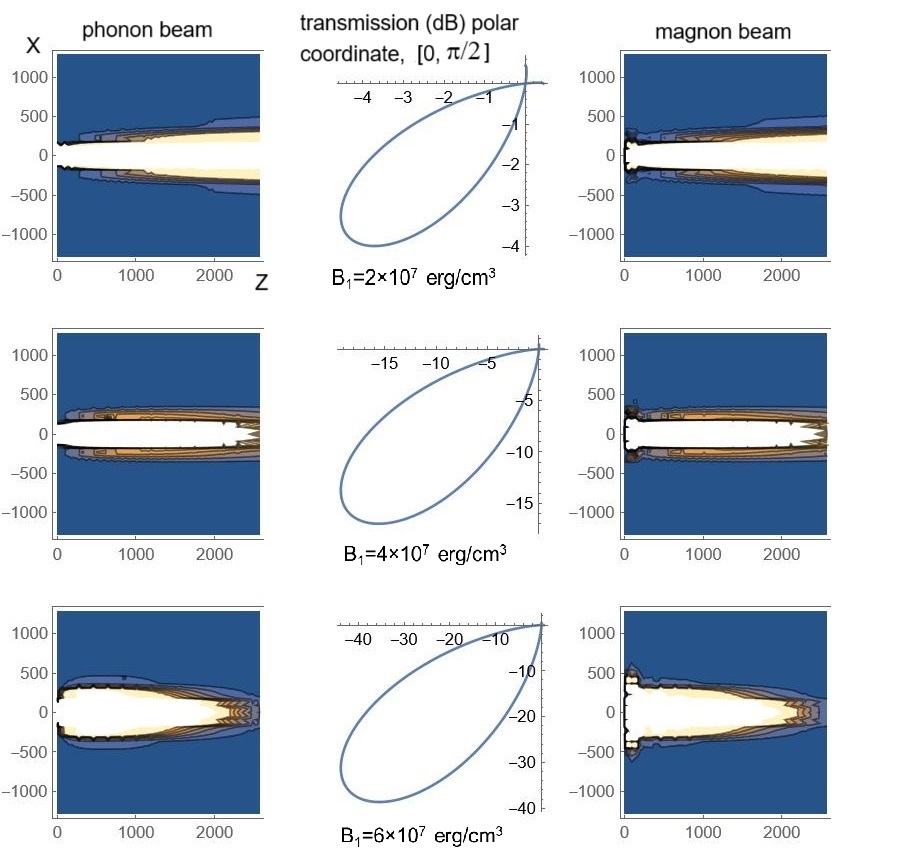}\caption{ Phonon (first column),
magnon (third column) beams at $\phi_{H}=\pi/4$, and the transmission signal
(second column) at different constants of magnetoelastic interaction (in one
row) and $H=50$ Oe, $v_{s}k/2\pi=287$ MHz. }%
\end{figure}It is easy to see that with the growth of the value of $B_{1}$ of
the magnetoelastic interaction, the turbulence in the beams of phonons and
magnons increases. At the same time, the attenuation of the transmitted signal
increases and it is maximum at an angle $\phi_{H}$ slightly less than $\pi/4$.
This fact agrees well with the observed data \cite{DBas19}-\cite{Focus}.

Let us now consider what the beams of phonons and magnons look like in
different magnetic fields. Figure 3 demonstrates that with increasing magnetic
field strength, the turbulent picture of the beams becomes calmer, and the
transmitted signal has less attenuation. This fact is understandable because
the phonons and magnons' "non-resonance" effect increases, effectively
reducing the magnetoelastic coupling.
\begin{figure}[ptbh]
\includegraphics[width=12cm]{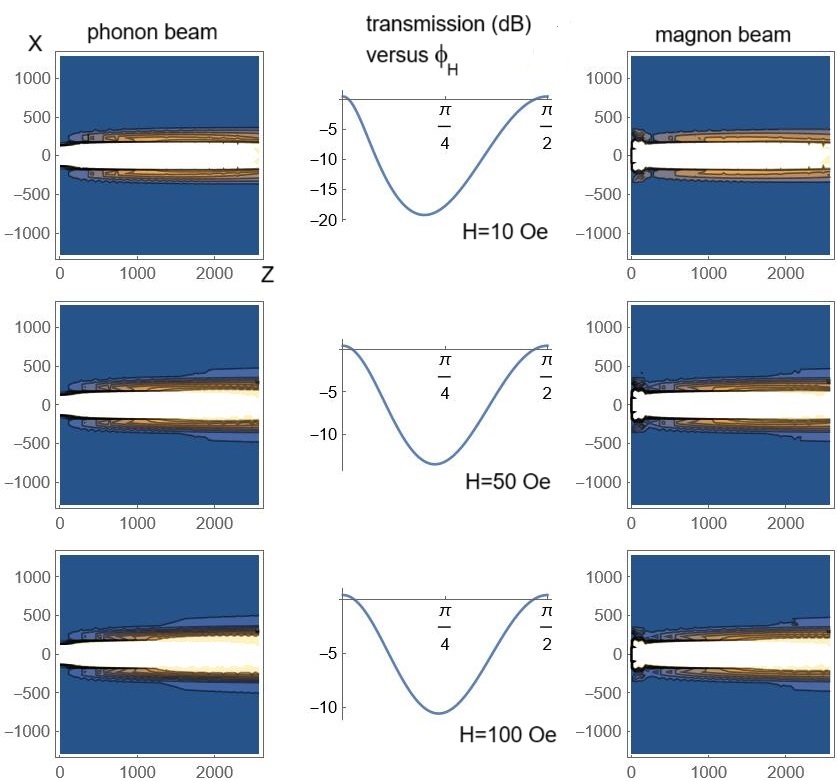}\caption{ Phonon beams (first column)
and magnon beams (third column) at $\phi_{H}=\pi/4$, and the angular
dependence of the transmission signal (second column) at different magnetic
fields (in one row). $B_{1}=3\times10^{7}$ erg/cm$^{3}$, $v_{s}k/2\pi=287$
MHz.}%
\end{figure}
\begin{figure}[ptbh]
\includegraphics[width=13cm]{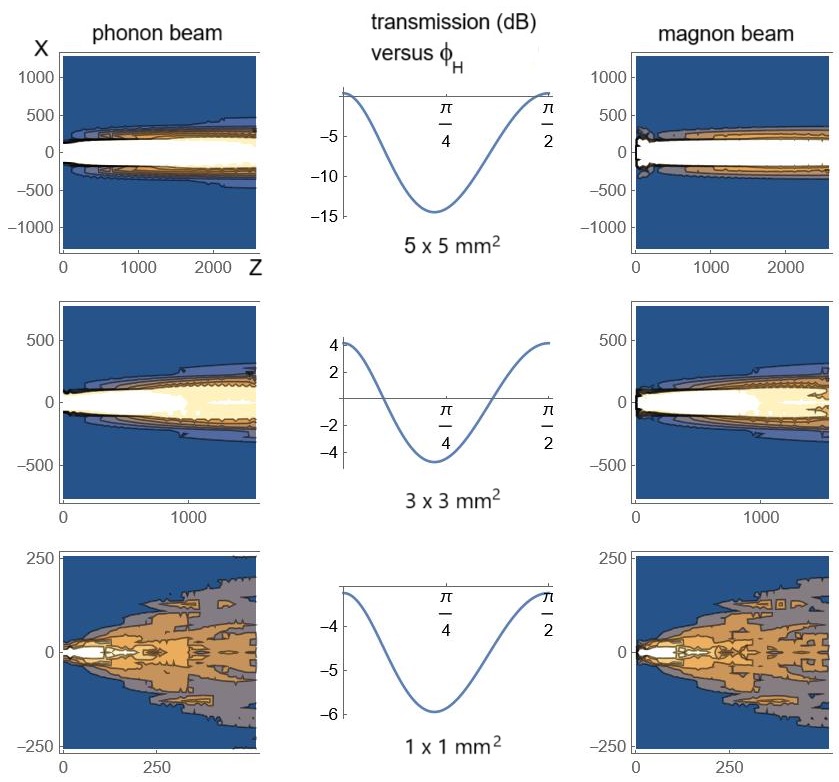}\caption{ Phonon (first column) and
magnon (third column) beams at $\phi_{H}=\pi/4$, and the angular dependence of
the transmission signal (second column) at different linear sizes of ultrathin
magnetic film (in one row). $H=50$ Oe, $B_{1}=3\times10^{7}$ erg/cm$^{3}$,
$v_{s}k/2\pi=287$ MHz.}%
\end{figure}Let us now consider how the beam pattern changes when the linear
dimensions of the ultrathin film change. Figure 4 shows the beams of phonons
and magnons and the angular dependence of the transmitted signal. On the one
hand, the phonon beam path becomes shorter, and thus, the attenuation becomes
more pronounced. On the other hand, magnons reflected from the upper and lower
film boundaries as their dimensions decrease can make an ever-increasing
contribution to the turbulence of both flows, which is probably what is
observed. 
%
\begin{figure}[ptbh]
\includegraphics[width=12cm]{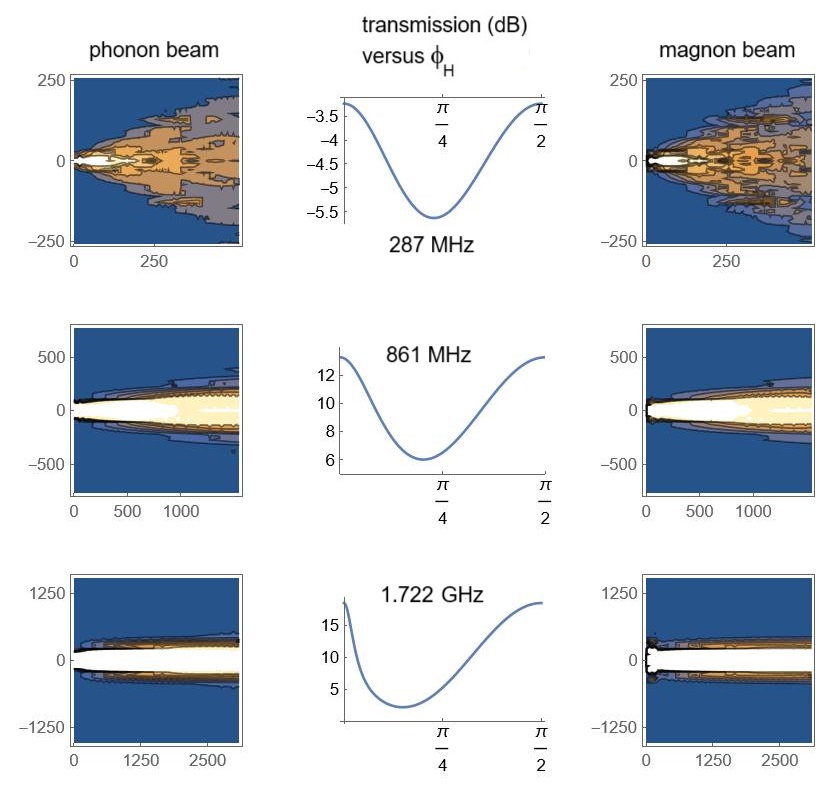}\caption{ Phonon (first column) and
magnon (third column) beams at $\phi_{H}=\pi/4$, and the angular dependence of
the transmission signal (second column) at different frequencies $v_{s}k/2\pi$
of an ultrathin magnetic film (in one row). $H=60$ Oe, $B_{1}=3\times10^{7}$
erg/cm$^{3}$, film $1\times1$ mm$^2$.}%
\end{figure}

The change in the turbulence of phonon and magnon beams with increasing
phonon-exciting frequency $v_{s}k_{e}/2\pi$ is shown in Fig. 5. We see that
the turbulence of both beams decreases with increasing excitation frequency.
This is due to the weakening of the role of magnons reflected from
the film boundaries with increasing $k_{e}$ (and the corresponding decrease in
wavelength). The increase in the deviation of the maximum absorption of the
phonon beam from the angle $\pi/4$ with increasing excitation frequency is noteworthy.

The analysis of the dependence of phonon and magnon beams on their
attenuation coefficients and on the film thickness showed that beam turbulence
(with a change in these coefficients by $\pm100\%$) has a weak dependence on them.
\section{Discussion}
In this paper, we considered a simple approach to describing unidirectional phonon and magnon beams coupled by magnetoelastic interaction in an ultrathin ferromagnetic film. In the weak turbulence approximation, equations (8)–(11) for the wave envelopes were obtained as a system of partial differential equations with nonlinear terms. Of course, this is a rich nonlinear system with various solutions depending on many external and internal parameters, possessing high predictive capabilities. Therefore, it makes perfect sense to conduct purposeful experiments in parallel with the proposed theory. As the above modeling showed, even in the linear approximation with non-resonant interaction of phonons and magnons with a given wave vector, we have non-trivial beam patterns that differ with changes in the magnetoelastic constant, magnetic field, film size, and phonon excitation frequency. Moreover, the obtained results describe the experimental results.

It is easy to guess that the equations for beams can also be obtained for
non-collinear cases when we have two or more sources of excitation of
quasiparticles in different directions. Naturally, the linear interactions of
phonons and magnons do not change the direction of the beams. New beams can
arise when three-particle and four-particle processes associated with
non-collinear excited beams are considered by momentum and energy
conservation laws. For example, from two beams of phonons with the same
frequency $\Omega$ and wave vectors $\boldsymbol{k}_{1}$ and $\boldsymbol{k}%
_{2}$, a beam of magnons with the wave vector $\boldsymbol{k=k}_{1}%
\boldsymbol{+k}_{2}$ and a forced oscillation frequency $2\Omega$ arises,
which does not necessarily coincide with the natural frequency $\omega
_{m\boldsymbol{k}}$ of these magnons. This example can clarify the effect of
phonon focusing observed experimentally.\cite{Focus}.

We are confident that the simple theory of beams of acoustic surface waves interacting with spin waves that we have developed will be of great interest to specialists studying and describing magnetoelastic dynamics in ultrathin magnetic films. Our model offers significant predictive capabilities of the theory and significantly simplifies calculations. In addition, we assume a broad interdisciplinary interest of specialists studying weak turbulence phenomena in condensed matter physics, plasma, and the physics of nonlinear wave processes in systems with two or more wave spectra.

\begin{acknowledgements}
This work is partially supported by the Air Force Office of Scientific Research (AFOSR) Award No. FA955023RXCOR001.
\end{acknowledgements}

\end{document}